\documentclass[prl,twocolumn,amsfonts]{revtex4}
\usepackage{graphicx}
\usepackage{amsmath}
\usepackage{amsfonts}
\usepackage[dvips]{epsfig}
\usepackage{bm}

\begin{document}
\author{Samuel Marcovitch and Benni Reznik }
\title{Is Communication Complexity Physical?}
\affiliation{ School of Physics and Astronomy,
Raymond and Beverly Sackler Faculty of Exact Sciences,
Tel-Aviv University, Tel-Aviv 69978, Israel.}

\date{\today}
\begin{abstract}
Recently, Brassard {\it{et. al.}} \cite{brassard} conjectured that
the fact that the maximal possible correlations between two
nonlocal parties are the quantum-mechanical ones
is linked to a reasonable restriction on communication complexity.
We provide further support for the conjecture in the multipartite case.
We show that any multipartite communication complexity problem
could be reduced to triviality, had Nature been more nonlocal than quantum-mechanics by a quite small gap for {\em any} number of parties.
Intriguingly, the multipartite nonlocal-box that we use to show the result corresponds to a generalized
Bell-Svetlichny inequality that manifests maximal genuine
multipartite nonseparability.
\end{abstract}
\maketitle

In a recent paper by Brassard {\it{et. al.}} \cite{brassard} a curious possibility regarding Nature was conjectured (BBLMTU conjecture):
the fact that the maximal possible correlations between two nonlocal parties are the quantum-mechanical (QM) ones
is linked to a reasonable restriction on communication complexity (CC).
Had the possible correlations been slightly higher,
CC would have been trivial.
BBLMTU use a hypothetical device, referred to as nonlocal-box (NLB),
in order to manifest their result.
NLBs have been initially suggested independently by Tsirelson \cite{cirelson}
and Popescu and Rohrlich \cite{pr}.
They are hypothetical devices which could be realized in causality respecting models of reality,
in which measures of nonlocality exceed the quantum-mechanical limit
and satisfy the maximal possible value.

The computational power of NLBs has been found independently by Van-Dam \cite{vandam} and Cleve \cite{buhrmancc}.
They have shown that equipped with many
NLBs two communicating parties
can reduce the required communication
between them to one bit,
{\it{i.e}} to trivial CC.
BBLMTU found that non-perfect NLBs suffice to reduce CC to triviality, given that we deal with probabilistic CC.
The bound they found for the probability of the NLB to operate is
$P_{CC}\approx 90.8\%$.
Quantum-mechanically, one can simulate NLBs with
probability $P_{QM}\approx 85.4\%$.
BBLMTU's bound is not necessarily the minimal one.
BBLMTU conjecture is that the actual bound is
$P_{CC}=P_{QM}+\epsilon$, where $\epsilon\rightarrow 0$.
The possible connection between communication complexity and physics
is further discussed by
Brassard \cite{brassard2} and Popescu \cite{pop2}.

In quantum-mechanics it is well known that nonlocality between many parties
is qualitatively different from two-party nonlocality.
If CC is indeed connected with physics, 
than the above conjecture should also be 
formulated and tested in the case of many communicating parties.
At first, it may seem that such a generalization cannot add new insights to
the problem, 
as it has been shown \cite{barrett} 
that {\em perfect} NLBs shared between any two parties of a multipartite system
are sufficient for computing any multipartite function with trivial CC.
However upon formulating the problem 
in terms of probabilistic CC
it can be shown that as the number of parties $N$ increases, 
the minimal required probability of the NLB to operate increases (subject to BBLMTU's method)
and departs still further from the QM bound.
Thus the conjecture seems to be refuted, 
unless new multipartite NLBs are considered. 

The main result of this paper is that any N-partite CC problem
can be reduced to triviality using a specific class of N-partite NLBs that operate
with probability $P_{CC}\approx 93.7\%$
for {\em any} $N$.
Quantum-mechanically too, the probability for successful 
simulation of the suggested NLB
remains constant as the number of parties increases,
$P_{QM}\approx 85.4\%$,
whereas a local-hidden-variable (LHV) theory shows decreasing probability
for increasing $N$, 
$$P_{LHV}=1/2+1/2^{\lfloor N/2\rfloor+1}.$$

In addition, we find an intriguing connection 
between this particular class of NLB and the violation of Bell's inequalities \cite{bell}.
Any multipartite NLB can be mapped to a generalized Bell's inequality (BI).
The BI that corresponds to our specific NLB is known as the generalized
Svetlichny's inequality \cite{svetlichny, svetlichny2, collins}.
This inequality measures genuine N-partite nonseparability.
We can define a theory of partial nonlocal correlations
(between any proper subset of the $N$ parties),
which maybe stronger than the QM correlations,
but still respect causality.
Svetlichny's inequality nevertheless gives 
a higher bound to the quantum-mechanical correlations. 
The generalized Svetlichny's inequality is exactly the one
we would expect the optimal multipartite
NLBs to correspond to since it implicitly relates to multipartite NLB correlations. 
Both the constant gap between trivial CC and QM, $P_{CC}-P_{QM}$, for every $N$,
as well as the correspondence between the suggested NLB and Svetlichny's inequality
seem to provide further support for BBLMTU conjecture.

To begin with, let us generalize communication complexity to the multipartite scenario.
Assume there is a boolean function $f$ of $N$ vectors of boolean arguments
$\{\boldsymbol{x}_1,\boldsymbol{x}_2,..,\boldsymbol{x}_N\}$,
where each of the N parties holds a different vector.
We are interested in the overall minimal number of bits required to
communicate between the parties so that $f$
can be computed.
In accordance with \cite{brassard},
we define distributive multipartite CC as a model
in which no communication is used.
Using only local operations each party produces a bit
$a_i$, $i=1,\dots,N$ such that $\Sigma_{i=1}^{N}a_i=f$, 
where throughout the paper sums are taken modulo $2$.
The probabilistic CC case requires that the sum equal $f$ with probability
$P(f)>0.5$, independent of the input size.
Clearly, if distributive multipartite CC is possible,
the communication required to compute $f$ 
reduces to triviality: 
only $k-1$ bits of communication over all parties.

Let us turn now to multipartite NLBs.
Assume a hypothetical model of reality in which
each of the parties inputs
a single bit $z_i$ to the NLB, where $1\leq i \leq N$.
The $N$-partite NLB fulfills
$\Sigma_{i=1}^N a_i=g(z_1,z_2,\dots,z_N)$,
where $g$ is an arbitrary multi-variable polynomial.
In the bipartite case there is a single distinct
NLB defined by 
\begin{equation}
\label{two}
a_1\oplus a_2=z_1\wedge z_2.
\end{equation}
In the multipartite case, however, one can define many ($2^N$) multipartite NLBs.
In the tripartite case, for example, one can define $g=z_1\wedge z_2\wedge z_3$ or
$g=z_1\wedge z_2\oplus z_1\wedge z_3 \oplus z_2\wedge z_3$.

We need a specific class of multipartite NLBs to reduce CC to triviality.
In order to find this, we now show the correspondence 
between multipartite NLBs and generalized BIs.
We will then derive the NLB corresponding to Svetlichny's inequality 
and find that this is just the class of NLB we have been looking for.
Generalizations of BIs to more than two parties
and more than two observables per party
have been extensively studied \cite{svetlichny,svetlichny2,collins,belinskii,mermin,ardehali,gisin,werner}.
We keep two observables per party and increase the number of parties.
In the bipartite case there is only one distinct measure, 
the Clauser-Horne-Shimony-Holt (CHSH) inequality \cite{chsh}:
\begin{equation}
A_2=C(x_1 x_2)+C(x_1 y_2)+C(y_1 x_2)-C(y_1 y_2)\leq2,
\end{equation}
where, for example, $C(x_1 x_2)$ is the correlation function of observable $x_1$ measured
by the first party and observable $x_2$ measured by the second party,
where each outcome equals $\pm 1$.
The generalized BIs that have corresponding NLBs are defined by
the subset in which all $2^N$ correlation functions appear and with $\pm1$ coefficients,
since correlations of $\pm1$ outcomes correspond to exclusive sums of boolean arguments,
as used in (\ref{two}).
For example, we map the bipartite NLB to CHSH inequality as follows:
inputs of NLB are $"0"$ or $"1"$, mapped to $"x"$ and $"y"$ observables respectively
for BI.
Outputs of NLB are $"0"$ and $"1"$, mapped to $"+1"$ and $"-1"$ respectively for BI.
The general process is straightforward:
$g$ has $2^N$ possible inputs,
which correspond to $2^N$ correlation functions in BI.
The outputs of the box are mapped to the corresponding correlations in BI.  
If the NLB operates properly, it maximizes the corresponding BI.
Thus, the probability that a theory can simulate the NLB 
is proportional to the corresponding measure of nonlocality (BI)
the theory holds.
Note that since there are doubly exponential ($2^{2^N}$)
generalizations of BI to the multipartite
case, most of the generalized BIs do not have corresponding multipartite NLBs.

We turn now to Svetlichny's inequality.
This is related to an inequality
with maximal QM violation
with respect to LHV theories,
suggested by Klyshko {\it{et. al}} \cite{belinskii}.
They showed that the N-partite inequality term $A_N$ can be recursively
constructed in the following way:
\begin{equation}
\label{m}
\begin{split}
& A_N\equiv \frac{1}{2}A_{N-1}\Big(x+y\Big)+\frac{1}{2}A'_{N-1}\Big(x-y\Big)\leq 2,
\\& A_2=xx+xy+yx-yy \leq 2,
\end{split}
\end{equation}
where $A'_N$ denotes the same expression as $A_N$ with all $"x"$ and $"y"$ exchanged
and $xx$, for example, denotes $C(x_1 x_2)$.
Note the normalization chosen for this definition.
Explicit computation will show that $A_N$ contains $2^N$ 
terms if N is even and $2^{N-1}$ terms if N is odd.
Therefore Klyshko's inequlaity has no corresponding NLB 
in the case where N is odd.
The generalized Svetlichny's inequality, however, which is defined as:
\begin{equation}
\label{svet}
S_N=\left\{ \begin{array}{ll}
A_N, & N\ \rm{even}\\
\frac{1}{2}(A_N\pm A_N'), & N\ \rm{odd}
\end{array} \right.
\end{equation}
has a corresponding multipartite NLB in all cases.

Let us now find the N-partite NLB that corresponds to $S_N$.
In general, we prove that by choosing a suitable 
mapping from Svetlichny's inequality and the NLB,
if there are $q$ $"1"$'s in the input of the NLB so that $q(q-1)/2$ is odd,
then the exclusive sum of the outputs of the box equals $1$.
Explicitly, the multipartite NLB that corresponds to (\ref{svet}) is
\begin{equation}
\label{cc}
\Sigma_{i=1}^N a_i=\Sigma_{i=1}^{N-1}\Sigma_{j=i+1}^N x_i \wedge x_j.
\end{equation}
The sum runs over all $N(N-1)/2$ pairs of the parties' inputs: $x_i$ and $x_j$.

Let us start with even $N$ for which Klyshko's 
inequality and Svetlichny's inequality coincide.
By induction, let us assume it holds for, say, $N=8n+4$
where we map the $"x"$ observable as the $"1"$ input, the $"y"$ 
observable as $"0"$ input, the $"+1"$ correlation as
$"1"$ output and the $"-1"$ correlation as $"0"$ output.
We can expand Klyshko's inequality:
\begin{equation}
\label{expand}
A_{N+2}= \frac{1}{2}A_N\Big(xy+yx\Big)+\frac{1}{2}A'_N\Big(xx-yy\Big)\leq 2.
\end{equation}
We choose for $A_{N+2}$ the same mapping as for $A_N$ with $"x"$ and $"y"$ replaced.
Positive sign correlations in $A_N$ have $p=(8n+4-q)$ $"y"$s,
therefore $p=1,2,5,6,9,10,\dots$.
From the first term in (\ref{expand}), we see that positive terms in $A_{8n+6}$ have
$p+1=2,3,6,7,10,11\dots$ $"y"$s, as required.
The second term in (\ref{expand}) involves $A'_N$,
in which the number of $"y"$s in positive sign correlations
is $q=2,3,6,7,10,11,\dots$.
Obviously, the addition of $"xx"$ adds no $"y"$s.
The number of $"y"$s in negative terms in $A'_N$ equals the
number of $"x"$s in negative terms in $A_N$, that is
$p=0,1,4,5,8,9,\dots$.
Adding two $"y"$s yields $p+2=2,3,6,7,10,11\dots$ $"y"$s, as required.
In the same manner, one can prove all other three cases: $N=8n$, $N=8n+2$ and $N=8n+6$.

For odd $N$ it can be recursively shown that (\ref{svet}) corresponds to (\ref{cc})
by choosing a suitable mapping 
and by taking the minus sign in (\ref{svet}) for $N=4n+3$ 
and the plus sign for $N=4n+1$,
where $n$ is integer. 
For example, for $N=3$ 
\begin{equation}
2S_3\!=\!xxx\!+\!xxy\!+\!xyx\!-\!xyy\!+\!yxx\!-\!yxy\!-\!yyx\!-\!yyy\! \leq \!2,
\nonumber
\end{equation}
which corresponds to (\ref{cc}) if we map the $"x"$ observable as the $"0"$ input, 
the $"y"$ observable as the $"1"$ input,
$"+1"$ correlation as $"0"$ output and $"-1"$ correlation as $"1"$ output.

The LHV-theory bound on $S_N$ is $2$, 
the QM bound is $2^{\frac{1}{2}+\lfloor \frac{N}{2}\rfloor}$
and the bound which corresponds to theories with partial correlations 
is $2^{\lfloor \frac{N}{2}\rfloor}$.
Note that these partial correlations 
are satisfied by the $N-1$-partite NLB that corresponds to
(\ref{cc}).

We would like now to sketch BBLMTU's method for achieving distributive
CC using non-perfect NLBs that operate
with probability $P_{CC}$.
Assume two parties compute an arbitrary function $f(\boldsymbol{x},\boldsymbol{y})$ 
distributively ($a\oplus b=f$) with probability $p>0.5$,
where one party posseses $\boldsymbol{x}$ and the other $\boldsymbol{y}$ 
and $p$ may depend on the size of the input
(such a probability always exists given shared randomness \cite{brassard}).
We would like to boost $p$ such that $f$ is computed distributively
with probability strictly higher than half, independent of the input size.
This is performed using a method originally suggested by von Neumann \cite{neumann},
by assuming the parties can compute nonlocal majority,
\begin{equation}
a\oplus b=\rm{Maj}(x_1\oplus y_1,x_2\oplus y_2,x_3\oplus y_3)
\end{equation} with probability $q$,
where $\rm{Maj}(u,v,w)$ equals the bit occurring most often among $u,v$ and $w$.
$p$ is boosted by computing $f$ distributively three times, and then computing the nonlocal majority of these three outcomes.
It can be shown that if $q>5/6$ and $1/2<p<s$, where
$\delta=q-5/6$ and $s=1/2+3\sqrt{\delta}/(2\sqrt{1+3\delta})$  
an iterative process of the above routine enables boosting $p$ to an arbitrary value
$1/2<t<s$.

Then BBLMTU show that nonlocal majority can be computed using two bipartite NLBs.
They first show that nonlocal equality:
\begin{equation}
a\oplus b=\left\{ \begin{array}{cc}
1 & if\ x_1\oplus y_1=x_2\oplus y_2=x_3\oplus y_3\\
0 & \rm{otherwise} \\
\end{array}\right\}
\end{equation}
can be computed using two bipartite NLBs.
Since $a=b$ is equivalent to $a\oplus \bar{b}$, we require
$$a\!\oplus \!b\!=\!(\!x'\!\oplus\!y'\!)\!\wedge\!(\!x''\!\oplus\!y''\!)\!=\!x'\!\wedge \!x''\!\oplus\!y''\!\wedge\! y''\! \oplus\! x'\!\wedge\! y'\! \oplus\! x''\!\wedge\! y'',$$
where $x'=x_1\oplus x_2$, $y'=y_1\oplus \bar{y_2}$ ,$x''=x_2\oplus x_3$, $y''=y_2\oplus \bar{y_3}$,
corresponding to local operations each party performs.
Recall that for the bipartite NLB, $a\oplus b=z_1\wedge z_2$.
We can immediately identify that nonlocal equality requires two NLBs:
$a'\oplus b'=x'\wedge y'$ and $a''\oplus b''=x''\wedge y''$,
such that the outputs bits of the two parties are
$a=(x'\wedge x'')\oplus a'\oplus a''$ and  $b=(y'\wedge y'')\oplus b'\oplus b''$.

Nonlocal majority can then be evaluated directly, by defining
$z=(\bar{a}\oplus b)+(z_1\oplus z_2\oplus z_3)$,
where $a$ and $b$ are the outputs of nonlocal equality and
$z_1=x_1\oplus y_1$, $z_2=x_2\oplus y_2$ and $z_3=x_3\oplus y_3$.
It is then straightforward to verify that $z$ equals nonlocal majority.
Now, since nonlocal majority should be computed with probability $q>5/6$ to boost the initial probability,
we require for NLB
$5/6<P_{CC}^2+(1-P_{CC})^2$
as the protocol given above succeeds precisely if none or both of the NLBs behave incorrectly.

We shall now generalize BBLMTU's method to the multipartite case.
We first note that corresponding to the bipartite case,
multipartite shared randomness enables computing $f$ distributively,
$\Sigma_{i=1}^N a_i=f(\boldsymbol{x}_1,\boldsymbol{x}_2,\dots,\boldsymbol{x}_N)$ 
with probability strictly higher than half (but dependent on the input size).
In accordance with BBLMTU we boost the probability using N-partite nonlocal majority:
\begin{equation}
\Sigma_{i=1}^N a_i=\rm{Maj}(\Sigma_{i=1}^N x^1_i,\Sigma_{i=1}^N x^2_i,\Sigma_{i=1}^N x^3_i),
\end{equation}
where $x^j_i$, denotes the j'th input of the i'th party.
Again, given that N-party nonlocal equality:
\begin{equation}
\Sigma_{i=1}^N a_i=(\Sigma_{i=1}^N x^1_i=\Sigma_{i=1}^N x^2_i)\wedge( \Sigma_{i=1}^N x^2_i=\Sigma_{i=1}^N x^3_i)
\end{equation}
is computed correctly, N-party nonlocal majority follows immediately:
define $z=(\bar{a_1}\oplus a_2\oplus \dots\oplus a_N)+(z^1\oplus z^2\oplus z^3),$
where $a_1,\ a_2,\dots, a_N$ are the distributed outputs of N-party nonlocal equality
and $z^j=\Sigma_{i=1}^N x^j_i$.

It remains, therefore, to find how many multipartite NLBs are required to calculate nonlocal equality.
Define $x'_i=x^1_i\oplus x^2_i$ for $1\leq i \leq N-1$ and $x'_N=x^1\oplus \bar{x}^2_N$ and
$x''_i=x^2_i\oplus x^3_i$ for $1\leq i \leq N-1$ and $x''_N=x^2\oplus \bar{x}^3_N$.
N-partite nonlocal equality can be recast as
\begin{equation}
\begin{split}
 \Sigma_{i=1}^N a_i&=(\Sigma_{i=1}^N x'_i )\wedge( \Sigma_{i=1}^N x''_i)
\\ &=\Sigma_{i=1}^N x'_i \wedge x''_i \oplus \Sigma_{i,j=1,i \neq j}^N x'_i \wedge x''_j.
\end{split}
\end{equation}
where the nonlocal result is expressed when $i\neq j$.

The N-partite NLB, which enables calculating N-partite nonlocal equality 
with the smallest number of NLBs, is exactly the one defined in (\ref{cc}). 
We require three such boxes to solve nonlocal equality for all N's.
Let each party $i$ enter $x'_i$ as input in the first box, $x''_i$ in the second box and $x'_i\oplus x''_i$ in the third box.
The outcomes of these boxes are correspondingly:
\begin{equation}
\begin{split}
 \Sigma_{i=1}^N a^1_i=&\Sigma_{i=1}^{N-1}\Sigma_{j=i+1}^N x'_i \wedge x'_j,
\\  \Sigma_{i=1}^N a^2_i=&\Sigma_{i=1}^{N-1}\Sigma_{j=i+1}^N x''_i \wedge x''_j,
\\  \Sigma_{i=1}^N a^3_i=&\Sigma_{i=1}^{N-1}\Sigma_{j=i+1}^N (x'_i\oplus x''_i) \wedge (x'_j \oplus x''_j)
\end{split}
\end{equation}
where $a^j_i$ denotes the j'th box output of the i'th player.
Since $a\oplus a=0$ for $a\in\{0,1\}$, the sum of outputs of all three boxes yields the required result:
$\Sigma_{i=1}^N a_i=\Sigma_{i,j=1,i \neq j}^N x'_i \wedge x''_j$.
The required probability for the specified NLB is therefore
$P_{CC}^3+3_{CC}(1-P_{CC})^2>5/6$, yielding $P_{CC}\approx 93.7\%$ for any number of parties.

We can immediately observe that $N(N-1)$ bipartite NLB are required to calculate N-partite nonlocal equality, subject to BBLMTU's method.
Use of multipartite NLBs is therefore significant because
we see that bipartite NLBs become less effective as the number
of parties increase.
Even without assuming any probability boosting technique, 
it is highly likely that the required probability of the bipartite 
NLB must increase as the number of parties increases.

As a final remark we would like to add that 
NLBs must also respect causality. 
Expressed in BI context, the probability
for any party to measure a certain outcome
should not depend on any other party's choice of 
which observable to measure (if any) and its outcome.
One can verify that a causality respecting model of 
NLB always exists in the multipartite case by
explicitly choosing the following set of probabilities \cite{prob}:
\begin{equation}
\begin{split}
&P(z_{i_1},z_{i_2},\dots,z_{i_k})=\frac{1}{2^k},
\\ &P(z_{i_1},z_{i_2},\dots,z_{i_N})=\frac{(-1)^t+s}{2^{N-1}},
\end{split}
\end{equation}
where $z_{i_j}$ corresponds to the outcome sign of the $"x"$ or $"y"$ measurements
of the $i_j$ party,
where $1\leq i_j\leq N$ and $1\leq k\leq N-1$.
$t$ is the number of $-1$ outcomes and
$s$ is the sign of the correlation in the 
generalized BI corresponding to the N-partite measurement.
That is, all joint probabilities over
less than N parties should be equally
spread and all joint probabilities over N
parties should equal zero or $2^{1-N}$ so that the corresponding
correlations equal plus or minus one, in accordance with the sign that appears
in the generalized BI.

In conclusion, we have examined BBLMTU conjecture in the broader sense of
multipartite properties of Nature. 
Multipartite nonlocality cannot be fully 
described in terms of bipartite nonlocality. 
Similarly, we showed that multipartite CC cannot be 
reduced to triviality using bipartite NLBs
that operate with a constant probability for any $N$,
subject to BBLMTU's method.
In the general $N$-partite case we provided
a bound for the maximal trivial CC -- QM gap needed to reduce CC to triviality.
Although this  bound is less tight compared with BBLMTU bound, the independence of
the number of parties seems to support CC's connection to physics in a wider context.
In addition, the correspondence between 
the optimal NLB for CC and Svetlichny's inequality, which measures genuine multipartite nonseparability,
suggests that this NLB truly generalizes multipartite nonlocality.
It provides stronger support for BBLMTU conjecture.

We thank J. Kupferman, N. Klinghoffer and M. Marcovitch for helpful discussions.
This work has been supported by the Israel Science Foundation grant number 784/06.


\begin{thebibliography}{99}
\bibitem{prob} It can be shown that this probability distribution is unique.
\end{thebibliography}
\end{document}